\documentclass{iopart}

\usepackage{graphicx}

\expandafter\let\csname equation*\endcsname\relax
\expandafter\let\csname endequation*\endcsname\relax

\usepackage{amsmath}
\usepackage{amsfonts}
\usepackage[dvips]{epsfig}
\usepackage{bm}

\usepackage[T1]{fontenc}
\usepackage[latin9]{inputenc}
\usepackage{float}
\usepackage{amsmath}
\usepackage{graphicx}
\usepackage{amssymb}


\begin{document}

\title{The Fermi Problem in Discrete Systems}

\author{Erez Zohar and Benni Reznik}
\address{School of Physics and Astronomy, Raymond and Beverly Sackler
Faculty of Exact Sciences, Tel-Aviv University, Tel-Aviv 69978, Israel.}

\begin{abstract}
The Fermi two-atom problem illustrates an apparent
causality violation in Quantum Field Theory which has to do with the nature of the built in correlations in the vacuum. It has been a constant subject of theoretical debate and discussions during the last few decades. Nevertheless, although the issues at hand could in principle be tested experimentally, the smallness of such apparent violations of causality in Quantum Electrodynamics prevented the observation of the predicted effect.
  In the present paper we show that the problem can be simulated within the framework of discrete systems that can be manifested, for instance, by trapped atoms in optical lattices or trapped ions.
Unlike the original continuum case, the causal structure  is no longer sharp.
 Nevertheless, as we show, it is possible to distinguish between "trivial" effects due to "direct" causality violations,  and the effects associated with Fermi's problem, even in such discrete settings. The ability to control externally the strength of the atom-field interactions, enables us also to study both the original Fermi problem with "bare atoms", as well as correction in the scenario that involves "dressed" atoms. Finally, we show that in principle, the Fermi effect can be detected using trapped ions.
\end{abstract}

\maketitle

\section{Introduction}

Fermi \cite{fermi} considered two atoms $A,B$, coupled to an electric
field. Consider The initial state $\left|\uparrow_{A}\downarrow_{B}0_{F}\right\rangle $ of the system
, where $\uparrow_{A}$ and $\downarrow_{B}$ denote atoms $A$ and $B$ in the excited and ground state respectivly, and $0_{F}$ - the ground state of the electric field. What is,
then, the probability of transition to the state $\left|\downarrow_{A}\uparrow_{B}0_{F}\right\rangle $?
Causal considerations lead us to believe that the probability must satisfy $P\varpropto\theta\left(t-\frac{R}{c}\right)$.
This is the result obtained by Fermi, perturbatively, by making a
further assumption, extending the domain of a frequency integration to include
negative frequencies as well. Shirokov, however, showed that the result is noncausal, and the causal probability was found by Fermi and others due to this approximation (see, for example, his review \cite{Shirokov}; Other method of calculation can be seen in \cite{power,rubin}).
The reason for that behaviour was examined by several authors (such as \cite{milonnijames,power,biswascompagno,cliche}).
A common solution to the problem is to reformulate the problem, stating that such a probability
is not measurable, since it requires a nonlocal instantaneous measurement
of both the atoms (and the field). Therefore some authors
did not post select the state of $A$ or of $A$ and the field
\footnote{Some Authors had even used a more general initial condition on $A$, or both $A$ and $B$.
}. It was shown in various ways, either perturbatively or exactly, that the transition
probability when only $B$ is measured is indeed causal, in the sense
that it is independent of atom $A$ for $t<R/c$ (where $R$ is the distance between the atoms), although it was shown that noncausal correlations develop between the ions (see, for example, \cite{biswascompagno}).

This leads us to the issue of \emph{dressing}:
It was shown \emph{mathematically} by Hegerfeldt \cite{hegerfeldt}
that within a renormalized theory, i.e. considering the dressed states
and Hamiltonian rather than the bare ones, the probability to find atom B excited (without post selecting A and the field) is not proportional
to $\theta\left(t-\frac{R}{c}\right)$.

It is therefore reasonable to believe that the existence of a state's
"tail", or of a "photon cloud" is the cause for what can
be interpreted as a noncausal result. Renormalization was introduced
by Hegerfeldt as a possible "way out"; The dressing process
of a two-level atom coupled to a field was investigated \cite{petrosky}: It was shown that both in
bare and dressed initial states of a single atom coupled to the field,
there is a nonlocal photon cloud surrounding the atom. In the bare
case it is canceled for $t=0$, and hence this nonlocal behaviour
is hidden in the initial condition ("curtain effect"), however,
for any positive time there is a photon distribution around the photon.

The Fermi Problem has been a constant subject of theoretical debate and discussions. Although the issues at hand could in principle be tested experimentally, the smallness of the apparent violations of causality in Quantum Electrodynamics prevented the observation of the basic effect as well as the study of the additional modifications due to the dressing of the photon cloud.

  The first objective of this article is to show that the Fermi problem can in fact be reformulated in the framework of discrete systems. This has the clear advantage as it enables to simulate the problem within the context of physical
         atomic systems such as trapped atoms in optical lattices \cite{ColdAtomBloch} or trapped ions \cite{james,singleion}. In another recent proposal, the continuous effect has been
          studied in the context of
         circuit QED \cite{Leon}.

However, the discreteness of our systems raises a new problem:   the concepts of "causality" and a "light-cone" now become ill defined and hence, non causal effect might obscure the effect we are looking for. We shall have a close look at such
errors and show how to distinguish between dynamical non-locality due to the finiteness
of the system and the Fermi Problem.

The use of atomic discrete systems has however many important advantages;
First, unlike QED, the use of vibrational degrees of freedom (phonons) as simulating the mediating field particles, and that
of the internal electronic levels as the "atom" detectors, allows for different time scales for
propagation and measurement, i.e. the measurement can be done at a much shorter time scale compared with the propagation time between the atoms.
Second, thanks to the developed quantum information techniques, the interaction between the
"atoms" and the "field" can be accurately controlled and hence turned on and off \cite{singleion}.
We will exploit this ability in order to control the dressing or the phonon cloud,
 or sometime to avoid the dressing and access bare states; something that in QED cannot be done.
 Finally, using quantum information techniques, it is possible to measure both the internal degrees of freedom of the "atoms" A and B, as well as the motional phonon state of the
  discrete field.
  Hence after turning the field-atom interaction off, before
the "causal time", one can perform, effectively, an instantaneous,
nonlocal measurement, by measuring the systems separately, which is
now possible since they are independent. Therefore one can also simulate the original model envisioned by Fermi, which assumes bare atoms and a final post selections of the atom $A$ and the field.

The article proceeds as follows: In section 2 we discuss the effect in discrete systems. After stating
the problem, we discuss the concepts of "causality" and an approximate
"lightcone" in such systems, which are different, of course,
than in the continuous case. We then present our study of the dressing process and discuss the possible scenarios in the presence of dressing.

In section 3 we show examples of the effect in two discrete
systems: the linear harmonic chain, and ions in a linear trap.

\section{The Effect in Discrete Systems}

\subsection{Statement of the Problem}

Consider a set of $N$ vibrating external degrees of freedom - atoms,
ions etc. We assume that the system has a quadratic Hamiltonian. The eigenmodes are created and annihilated by $a_{k}^{\dagger},a_{k}$,  ($\left[a_{k},a_{l}^{\dagger}\right]=\delta_{kl}$), and the eigenmodes' frequencies are $\left\{ \omega_{k}\right\} $.
Hence, the displacement (from equilibrium) operators have the general
form\begin{equation}
q_{n}=\overset{N-1}{\underset{k=0}{\sum}}\left(\lambda_{nk}a_{k}+\lambda_{nk}^{*}a_{k}^{\dagger}\right)\end{equation}
and the Hamiltonian is ($\hbar=1$) \begin{equation}
H_{F}=\overset{N-1}{\underset{k=0}{\sum}}\omega_{k}a_{k}^{\dagger}a_{k}\label{eq:HF}\end{equation}

Besides the external degrees of freedom, we assume two-level internal
degree of freedom on each site, represented by a spin-1/2
system, with the free Hamiltonian\begin{equation}
H_{0}^{n}=\frac{1}{2}\Omega_{n}\sigma_{z}^{n}\end{equation}
which interacts with the external (vibrational) degree of freedom
in the same site in the following local manner:

\begin{equation}
H_{I}^{n}=\epsilon f_{n}\left(t\right)q_{n}\sigma_{x}^{n}.\end{equation}

The Fermi Problem and other related problems where atoms are coupled
to a field were mostly formulated with the atoms coupled to an electric
field $\mathbf{E}$, but also to a scalar field, where the coupling
is either to its conjugate momentum (such as in \cite{petrosky})
or to the field operator itself (such as in \cite{cliche}).
The 'noncausal' effect should not differ across these cases, since field
propagators are not causal, either in scalar or electromagnetic field theory. Therefore, for the sake of simplicity, we shall work
in the scalar case, where the atoms are coupled to the field. Moreover, this is the interaction used in the case of ion traps, as in the relevant section in this paper, and hence it is used throughout other parts of the paper as well.

The total Hamiltonian is then: \begin{equation}
H=H_{F}+\overset{N-1}{\underset{n=0}{\sum}}\left(H_{0}^{n}+H_{I}^{n}\right)\end{equation}

Note that this Hamiltonian can be implemented in various ways, such
as using ion traps (\cite{james,singleion}) and optical lattices (\cite{ColdAtomBloch,singleaddressing}). One of the main
requirements from the system is the possibility to measure the internal
degree of freedom fast enough.

After defining the physical systems of interest, we can state Fermi's
Problem. The external motional degrees of freedom will
serve as a discrete analog to the field. We pick two sites $A$ and $B$ to
act as the atoms which carry internal
levels. Therefore we set, for every $n\notin\left\{ A,B\right\} $,
$f_{n}\left(t\right)\equiv0$.

The system is prepared in the initial
state\begin{equation}
\left|\psi\left(t=0\right)\right\rangle \equiv\left|\psi_{i}\right\rangle =\left|\uparrow_{A}\downarrow_{B}0_{F}\right\rangle \label{eq:InSt}\end{equation}
and the transition probability $P\left(t\right)$ to the final state\begin{equation}
\left|\psi_{f}\right\rangle =\left|\downarrow_{A}\uparrow_{B}0_{F}\right\rangle \label{eq:FiSt}\end{equation}
is examined.

As in Fermi's original formulation, after turning the interactions
off, the state of all three systems ($A,B,F$) can be measured separately,
in an effectively instantaneous-nonlocal measurement, which is not
possible if the interactions are constantly open. Therefore this post-selection,
which was "forbidden", is possible, since it can be performed after the
interactions are switched off.

\subsection{Causality}
We shall develop here a method for distinguishing the a-causal contribution that has a
dynamical source, from the a-causal contributions that arise in the Fermi Problem.
Let us calculate the transition amplitude using the interaction-picture, with the interaction
Hamiltonians (for $A,B$) \begin{equation}
H_{I}^{n}=\epsilon f_{n}\left(t\right)\left(\sigma_{+}^{n}e^{i\Omega_{n}t}+\sigma_{-}^{n}e^{-i\Omega_{n}t}\right)\underset{k}{\sum}\left(\lambda_{nk}a_{k}e^{-i\omega_{k}t}+\lambda_{nk}^{*}a_{k}^{\dagger e^{i\omega_{k}t}}\right)\label{eq:Hint}\end{equation}

The $\left|\psi_{i}\right\rangle \longrightarrow\left|\psi_{f}\right\rangle $
transition amplitude is (see appendix 1 for details) can be written (in leading order) as \begin{multline}
-\frac{A\left(t\right)}{\epsilon^{2}}=\frac{1}{2}\int_{0}^{t}dt'f_{A}\left(t'\right)e^{-i\Omega_{A}t'}\int_{0}^{t}dt''f_{B}\left(t''\right)e^{i\Omega_{B}t''}\left\langle 0\right|\left\{ q_{A}\left(t'\right),q_{B}\left(t''\right)\right\} \left|0\right\rangle +\\
\\\frac{1}{2}\int_{0}^{t}dt'f_{A}\left(t'\right)e^{-i\Omega_{A}t'}(\int_{0}^{t'}dt''f_{B}\left(t''\right)e^{i\Omega_{B}t''}\left\langle 0\right|\left[q_{A}\left(t'\right),q_{B}\left(t''\right)\right]\left|0\right\rangle -\\
\int_{t'}^{t}dt''f_{B}\left(t''\right)e^{i\Omega_{B}t''}\left\langle 0\right|\left[q_{A}\left(t'\right),q_{B}\left(t''\right)\right]\left|0\right\rangle )\label{eq:Amplitude}\end{multline}
Hence it decomposes to two parts: the first is commutative, correlation-dependent
and contains two independent time integrations,\begin{equation}
-\frac{A_{0}\left(t\right)}{\epsilon^{2}}=\frac{1}{2}\int_{0}^{t}dt'f_{A}\left(t'\right)e^{-i\Omega_{A}t'}\int_{0}^{t}dt''f_{B}\left(t''\right)e^{i\Omega_{B}t''}\left\langle 0\right|\left\{ q_{A}\left(t'\right),q_{B}\left(t''\right)\right\} \left|0\right\rangle \end{equation}
and the second is non-commutative, commutator-dependent and integration
order dependent,\begin{multline}
-\frac{A_{c}\left(t\right)}{\epsilon^{2}}=\\
\frac{1}{2}\int_{0}^{t}dt'f_{A}\left(t'\right)e^{-i\Omega_{A}t'}(\int_{0}^{t'}dt''f_{B}\left(t''\right)e^{i\Omega_{B}t''}\left\langle 0\right|\left[q_{A}\left(t'\right),q_{B}\left(t''\right)\right]\left|0\right\rangle -\\
\int_{t'}^{t}dt''f_{B}\left(t''\right)e^{i\Omega_{B}t''}\left\langle 0\right|\left[q_{A}\left(t'\right),q_{B}\left(t''\right)\right]\left|0\right\rangle )\end{multline}

In the continuum field limit, $q_{n}\left(t\right)\longrightarrow\phi\left(x_{n},t\right)$,
the commutator vanishes for a spacelike separation ($t<\frac{|x_{B}-x_{A}|}{c}$)
and therefore only the contribution of $A_{0}\left(t\right)$ remains.
This is the part which dominates and "causes" the nonlocal Fermi effect;
Therefore, we would like to show that for discrete systems, where
$A_{c}\left(t\right)$ is nonvanishing, its contribution is much smaller
or negligible compared with  $A_{0}\left(t\right)$, for times
that are restricted within the effective lightcone.

Generally,\begin{equation}
q_{n}\left(t\right)=\underset{k}{\sum}\left(\lambda_{nk}a_{k}e^{-i\omega_{k}t}+\lambda_{nk}^{*}a_{k}^{\dagger}e^{i\omega_{k}t}\right)\end{equation}
and hence, defining $\tau=t''-t'$, \begin{equation}
\left\langle 0\right|q_{A}\left(t'\right)q_{B}\left(t''\right)\left|0\right\rangle =\underset{k}{\sum}\left\langle 0\right|\lambda_{Ak}\lambda_{Bk}^{*}a_{k}a_{k}^{\dagger}\left|0\right\rangle =\underset{k}{\sum}\lambda_{Ak}\lambda_{Bk}^{*}e^{i\omega_{k}\tau}\end{equation}

Define the functions\begin{equation}
F_{a}\left(\tau\right)\equiv\left\langle 0\right|\left\{ q_{A}\left(t'\right),q_{B}\left(t''\right)\right\} \left|0\right\rangle =\underset{k}{\sum}\left(\lambda_{Ak}\lambda_{Bk}^{*}e^{i\omega_{k}\tau}+\lambda_{Ak}^{*}\lambda_{Bk}e^{-i\omega_{k}\tau}\right)\label{Fa}\end{equation}
and\begin{equation}
iF_{c}\left(\tau\right)\equiv\left\langle 0\right|\left[q_{A}\left(t'\right),q_{B}\left(t''\right)\right]\left|0\right\rangle =\underset{k}{\sum}\left(\lambda_{Ak}\lambda_{Bk}^{*}e^{i\omega_{k}\tau}-\lambda_{Ak}^{*}\lambda_{Bk}e^{-i\omega_{k}\tau}\right)\label{Fc}\end{equation}

In the continuum, $F_{c}\left(\tau\right)\sim\theta\left(\tau-\frac{|x_{B}-x_{A}|}{c}\right)$;
So the goal is to show that in the relevant discrete systems,
there is a typical velocity $c$, for which $F_{c}\left(\tau\right)$
is negligible , and is particularly small compared to $F_{a}\left(\tau\right)$,
for times satisfying $\tau<\frac{|x_{B}-x_{A}|}{c}$.

\subsection{Dressing}

The difference between bare and dressed states plays a major role
in the Fermi problem, essentially because when the interaction between
the field and the atoms is constantly open, one should not consider
the bare states, but rather the "true" states which are the
dressed ones. Hegerfeldt, in his theorem \cite{hegerfeldt}, referred
to the states as "renormalized" ones, and suggested the renormalization
- the presence of a photon cloud around each atom - as the cause of
what appears as a noncausal result. We interpret the "renormalization" - i.e., the process of obtaining the "physical" states of the system - as dressing (unlike the meaning of renormalization in QFT) and perform it perturbatively, as in \cite{petrosky,Dressing}.

In the cases discussed here, however, dressing is not mandatory. As
remarked earlier in this paper, it depends on the opening manner
of the interaction functions $f_{n}\left(t\right)$ of $A$ and $B$.

Another related issue is the field excitations distribution, which is discussed in appendix 2.

Consider two different opening schemes:
\begin{enumerate}
\item The opening functions satisfy $f_{A,B}\left(t\right)\neq0$ only for
a very short period of time, $T$. If we set $T$ to be smaller than
the time of propagation between the two sites, we can be sure that
positive probability is due to the effect. We can also choose
$T$ to be small enough so that dressing effects become irrelevant.
The physical process can then be described with bare states.

\item Turn the interactions on adiabatically, causing the system to dress.
Then, at $t=0$, bringing them to the desired \emph{dressed} initial
state, and allowing the system to evolve until $t=T$, when
the interactions are set off. As in the previous scheme, $T$ can be chosen to be smaller
than the causal time.\end{enumerate}
 After turning the interactions off we have
here too the benefit of effectively carrying out a nonlocal measurement.
Let us discuss the second opening scheme in detail.

\subsubsection{Ground State Dressing}

Suppose we use interaction functions of the form\begin{equation}
f_{A}\left(t\right)=f_{B}\left(t\right)=\begin{cases}
e^{\frac{t}{\tau}} & t<0\\
f_{0}\left(t\right) & 0\leq t\leq T\\
0 & t\geq0\end{cases}\end{equation}
we start, at $t\longrightarrow-\infty$ with the initial bare ground
state \begin{equation}
\left|\psi\left(t\longrightarrow-\infty\right)\right\rangle \equiv\left|G^{\left(0\right)}\right\rangle =\left|\downarrow_{A}\downarrow_{B}0_{F}\right\rangle \end{equation}
Assuming, as before, that $\epsilon$ is small enough, we can calculate
the time evolution in time-dependent perturbation theory. We take $\tau$ to be large enough, so the interaction is
turned on adiabatically, and thus according to the adiabatic theorem,
\begin{equation}
\left|\psi\left(t=0\right)\right\rangle \equiv\left|G\right\rangle =\left|G^{\left(0\right)}\right\rangle +\epsilon\left|G^{\left(1\right)}\right\rangle +\epsilon^{2}\left|G^{\left(2\right)}\right\rangle +O\left(\epsilon^{3}\right)\end{equation}
where the corrections $\left|G^{\left(n\right)}\right\rangle $ are
calculated using time-independent perturbation theory. This is the
dressed ground state.

The first order correction
\footnote{For simplicity, it is assumed here that $\Omega_{A}=\Omega_{B}\equiv\Omega$.} is\begin{equation}
\left|G^{\left(1\right)}\right\rangle =-\underset{k}{\sum}\frac{\lambda_{Ak}^{*}\left|\uparrow_{A}\downarrow_{B}1_{kF}\right\rangle +\lambda_{Bk}^{*}\left|\downarrow_{A}\uparrow_{B}1_{kF}\right\rangle }{\Omega+\omega_{k}}\end{equation}
As a first order expression, it results from one operation of the
interaction Hamiltonian, and therefore contains (as can be seen)
only separate contributions from interactions with either $A$ or $B$. Therefore, in
the first order, the correction to the ground state of two sites coupled
to the field is the same as coupling each of them separately.

The second order correction can be decomposed into two parts:\begin{equation}
\left|G^{\left(2\right)}\right\rangle =\left|G^{\left(2,1\right)}\right\rangle +\left|G^{\left(2,2\right)}\right\rangle \end{equation}
where\begin{equation}
\left|G^{\left(2,1\right)}\right\rangle =\underset{k}{\sum}\frac{\left(\lambda_{Ak}^{*}\right)^{2}+\left(\lambda_{Bk}^{*}\right)^{2}}{\sqrt{2}\left(\Omega+\omega_{k}\right)\omega_{k}}\left|\downarrow_{A}\downarrow_{B}2_{kF}\right\rangle +\underset{k,l\neq k}{\sum}\frac{\lambda_{Ak}^{*}\lambda_{Al}^{*}+\lambda_{Bk}^{*}\lambda_{Bl}^{*}}{\left(\Omega+\omega_{k}\right)\left(\omega_{k}+\omega_{l}\right)}\left|\downarrow_{A}\downarrow_{B}1_{k}1_{lF}\right\rangle \end{equation}
is due to separate dressing processes ($A,B$ alone), and \begin{multline}
\left|G^{\left(2,2\right)}\right\rangle =\underset{k}{\sum}\frac{\lambda_{Ak}^{*}\lambda_{Bk}+\lambda_{Bk}^{*}\lambda_{Ak}}{2\Omega\left(\Omega+\omega_{k}\right)}\left|\uparrow_{A}\uparrow_{B}0_{F}\right\rangle +\underset{k}{\sum}\frac{\sqrt{2}\lambda_{Ak}^{*}\lambda_{Bk}^{*}}{\left(\Omega+\omega_{k}\right)^{2}}\left|\uparrow_{A}\uparrow_{B}2_{kF}\right\rangle +\\
\underset{k,l\neq k}{\sum}\frac{\lambda_{Ak}^{*}\lambda_{Bl}^{*}+\lambda_{Al}^{*}\lambda_{Bk}^{*}}{\left(\Omega+\omega_{k}\right)\left(2\Omega+\omega_{k}+\omega_{l}\right)}\left|\uparrow_{A}\uparrow_{B}1_{k}1_{lF}\right\rangle \end{multline}
is the result of dressing caused by interactions with both $A$ and $B$ together.

In general, odd orders of the perturbative series have states with
total spin (in the z direction) zero, and an odd number of phonons, whereas
even orders have $\pm1$ total spin and an even number of phonons.

\subsubsection{Excitation of $A$ and Selection of the Initial Dressed State}

The bare processes started with the initial state $\left|\psi_{i}^{\left(0\right)}\right\rangle =\left|\uparrow_{A}\downarrow_{B}0_{F}\right\rangle =\sigma_{x}^{A}\left|G^{\left(0\right)}\right\rangle $.
What is the equivalent dressed state? In order to do that, let us
operate with $\sigma_{x}^{A}$ on $\left|G\right\rangle $.

This is implemented using the approximation of an impulsive interaction, $H'=\alpha\delta\left(t\right)\sigma_{x}^{A}$. Therefore, the evolution of the system neglecting the "free" parts is \begin{equation}
\left|\psi\left(t=0+\right)\right\rangle =e^{-i\alpha\sigma_{x}^{A}}\left|G\right\rangle \end{equation}

choosing $\alpha=\pi/2$, the state is found to be (up to an irrelevant global phase)\begin{equation}
\left|\psi\left(t=0+\right)\right\rangle =\sigma_{x}^{A}\left|G\right\rangle \end{equation}
hence we get that the "initial state", i.e. the state at $t=0$
is\begin{equation}
\left|\psi_{i}\right\rangle =\left|\psi_{i}^{\left(0\right)}\right\rangle +\epsilon\left|\psi_{i}^{\left(1\right)}\right\rangle +\epsilon^{2}\left|\psi_{i}^{\left(2,1\right)}\right\rangle +\epsilon^{2}\left|\psi_{i}^{\left(2,2\right)}\right\rangle +O\left(\epsilon^{3}\right)\end{equation}
where the corrections to the bare initial state $\left|\psi_{i}^{\left(0\right)}\right\rangle =\left|\uparrow_{A}\downarrow_{B}0_{F}\right\rangle $
are\begin{equation}
\left|\psi_{i}^{\left(1\right)}\right\rangle =-\underset{k}{\sum}\frac{\lambda_{Ak}^{*}\left|\downarrow_{A}\downarrow_{B}1_{kF}\right\rangle +\lambda_{Bk}^{*}\left|\uparrow_{A}\uparrow_{B}1_{kF}\right\rangle }{\Omega+\omega_{k}}\end{equation}
and\begin{equation}
\left|\psi_{i}^{\left(2,1\right)}\right\rangle =\underset{k}{\sum}\frac{\left(\lambda_{Ak}^{*}\right)^{2}+\left(\lambda_{Bk}^{*}\right)^{2}}{\sqrt{2}\left(\Omega+\omega_{k}\right)\omega_{k}}\left|\uparrow_{A}\downarrow_{B}2_{kF}\right\rangle +\underset{k,l\neq k}{\sum}\frac{\lambda_{Ak}^{*}\lambda_{Al}^{*}+\lambda_{Bk}^{*}\lambda_{Bl}^{*}}{\left(\Omega+\omega_{k}\right)\left(\omega_{k}+\omega_{l}\right)}\left|\uparrow_{A}\downarrow_{B}1_{k}1_{lF}\right\rangle \end{equation}
\begin{multline}
\left|\psi_{i}^{\left(2,2\right)}\right\rangle =\underset{k}{\sum}\frac{\lambda_{Ak}^{*}\lambda_{Bk}+\lambda_{Bk}^{*}\lambda_{Ak}}{2\Omega\left(\Omega+\omega_{k}\right)}\left|\downarrow_{A}\uparrow_{B}0_{F}\right\rangle +\underset{k}{\sum}\frac{\sqrt{2}\lambda_{Ak}^{*}\lambda_{Bk}^{*}}{\left(\Omega+\omega_{k}\right)^{2}}\left|\downarrow_{A}\uparrow_{B}2_{kF}\right\rangle +\\
\underset{k,l\neq k}{\sum}\frac{\lambda_{Ak}^{*}\lambda_{Bl}^{*}+\lambda_{Al}^{*}\lambda_{Bk}^{*}}{\left(\Omega+\omega_{k}\right)\left(2\Omega+\omega_{k}+\omega_{l}\right)}\left|\downarrow_{A}\uparrow_{B}1_{k}1_{lF}\right\rangle \label{eq:dressed}\end{multline}
what can immediately seen from this result, is that the desired final
state $\left|\downarrow_{A}\uparrow_{B}0_{F}\right\rangle $
\footnote{We measure the bare states, since the measurement is assumed to take
a very short time and is therefore unable to distinguish between the
bare and dressed energies.
} is already contained in the dressed initial superposition.

The next question is what is the correct approximation of the dressed
initial state. There are several ways to answer this question:
\begin{enumerate}
\item Using $\left|\psi_{i}\right\rangle =\sigma_{x}^{A}\left|G\right\rangle $
as an approximation for the dressed initial state and starting the
time evolution. As mentioned before, in that case the probability
we seek will not start from zero, since the final state already appears
in the initial superposition.
\item Measuring $A$'s spin straightly after applying $\sigma_{x}^{A}$,
and selecting only the up spins - which is like operating with $\sigma_{+}^{A}$
instead; This should apply if one claims that $\left|\psi_{i}^{\left(0\right)}\right\rangle =\sigma_{+}^{A}\left|G^{\left(0\right)}\right\rangle $
rather than $\left|\psi_{i}^{\left(0\right)}\right\rangle =\sigma_{x}^{A}\left|G^{\left(0\right)}\right\rangle $
and therefore the same should be done with the dressed state. In this
case, the effects of mutual dressing are lost (in the leading order)
and hence the final state does not appear in the initial superposition
and the probability starts from zero. This is the scheme used in \cite{petrosky}.
\item Returning to the bare state: Measuring both $A$ and $B$'s spins,
projecting to the bare initial state $\left|\uparrow_{A}\downarrow_{B}0_{F}\right\rangle $.
This, of course, "ruins" the dressing process - however it is
good for comparison.
\end{enumerate}
Finally, we note that in the dressed cases (the first two) the dressed states obtained perturbatively are not normalized. In
order to normalize them (to order $\epsilon^{2}$) one should add $-\frac{1}{2}\epsilon^{2}\left\Vert \psi_{i}^{\left(1\right)}\right\Vert ^{2}\left|\psi_{i}^{\left(0\right)}\right\rangle $.
This additional term does not contribute to the calculations and processes discussed in the following, and hence we ignore it.

\subsubsection{Time Evolution of the Dressed State}

After selecting the initial state, it evolves until $t=T$ similarly
to the bare state. The time evolution of the dressed state is (we define $H_{I}=\epsilon V$):\begin{multline}
\left|\psi_{i}\left(t\right)\right\rangle =\left|\psi_{i}^{\left(0\right)}\right\rangle +\epsilon\left(-i\int_{0}^{t}dt'V\left(t'\right)\left|\psi_{i}^{\left(0\right)}\right\rangle +\left|\psi_{i}^{\left(1\right)}\right\rangle \right)+\\
\epsilon^{2}\left(-\int_{0}^{t}dt'V\left(t'\right)\int_{0}^{t'}dt''V\left(t''\right)\left|\psi_{i}^{\left(0\right)}\right\rangle -i\int_{0}^{t}dt'V\left(t'\right)\left|\psi_{i}^{\left(1\right)}\right\rangle +\left|\psi_{i}^{\left(2\right)}\right\rangle \right)+O\left(\epsilon^{3}\right)\label{eq:DressedTimeEvolution}\end{multline}

The effect's amplitude is then (in leading order)\begin{equation}
A\left(t\right)=\epsilon^{2}\underset{k}{\sum}\left(\lambda_{Bk}^{*}\lambda_{Ak}F_{1k}\left(t\right)+\lambda_{Ak}^{*}\lambda_{Bk}F_{2k}\left(t\right)\right)\label{eq:ADressed}\end{equation}
where\begin{multline}
F_{1k}\left(t\right)=-\int_{0}^{t}dt'f_{0}\left(t'\right)e^{-i\left(\Omega+\omega_{k}\right)t'}\int_{0}^{t'}dt''f_{0}\left(t''\right)e^{i\left(\Omega+\omega_{k}\right)t''}\\
+\frac{i(d_{1}+d_{2})}{\Omega+\omega_{k}}\int_{0}^{t}dt'f_{0}\left(t'\right)e^{-i\left(\Omega+\omega_{k}\right)t'}+\frac{d_{1}}{2\Omega\left(\Omega+\omega_{k}\right)}\label{eq:FDressed}\end{multline}
\begin{multline}
F_{2k}\left(t\right)=-\int_{0}^{t}dt'f_{0}\left(t'\right)e^{i\left(\Omega-\omega_{k}\right)t'}\int_{0}^{t'}dt''f_{0}\left(t''\right)e^{-i\left(\Omega-\omega_{k}\right)t''}\\
+\frac{id_{1}}{\Omega+\omega_{k}}\int_{0}^{t}dt'f_{0}\left(t'\right)e^{i\left(\Omega-\omega_{k}\right)t'}+\frac{d_{1}}{2\Omega\left(\Omega+\omega_{k}\right)}\label{eq:FDressed2}\end{multline}
and the constants $d_{1},d_{2}$ set the dressing (or non-dressing) scheme,
as defined in the previous section: scheme 1 ($\sigma_{x}^{A}$) corresponds
to $d_{1}=1,d_{2}=0$, scheme 2 ($\sigma_{+}^{A}$) to $d_{1}=0,d_{2}=1$, and scheme 3
(no dressing) to $d_{1}=d_{2}=0$.

\section{Examples}

\subsection{Harmonic Chain}

As a first example, consider a linear harmonic chain with nearest-neighbour interactions.

\subsubsection{Properties of the Harmonic Chain}

Consider a circular (periodic) linear chain of $N$ masses $M$, coupled by springs whose constant is $K$, described by the Hamiltonian\begin{equation}
H_{F}=\frac{1}{2}\underset{n=1}{\overset{N}{\sum}}\left(\frac{\pi_{n}^{2}}{M}+M\nu^{2}\xi_{n}^{2}+K\left(\xi_{n}-\xi_{n-1}\right)^{2}\right)\end{equation}

As in \cite{reznikbotero}, we perform the canonical transformation\begin{equation}
q_{n}=\sqrt{M\nu\sqrt{1+\frac{2K}{M\nu^{2}}}}\xi_{n}\end{equation}
\begin{equation}
p_{n}=\frac{\pi_{n}}{\sqrt{M\nu\sqrt{1+\frac{2K}{M\nu^{2}}}}}\end{equation}
and set\begin{equation}
E_{0}=\nu\sqrt{1+\frac{2K}{M\nu^{2}}}\end{equation}
\begin{equation}
\alpha=\frac{\frac{2K}{M\nu^{2}}}{1+\frac{2K}{M\nu^{2}}}\label{eq:alpha}\end{equation}
resulting in the Hamiltonian\begin{equation}
H_{F}=\frac{E_{0}}{2}\overset{N-1}{\underset{n=0}{\sum}}\left(p_{n}^{2}+q_{n}^{2}-\alpha q_{n}q_{n+1}\right)\end{equation}

Equation {[}\ref{eq:alpha}{]} implies $0<\alpha<1$. $\alpha\rightarrow1$
is the strong coupling limit, where $\frac{2K}{M\nu^{2}}\rightarrow\infty$.

Periodic boundary conditions are imposed, and thus the equations of
motion,\begin{equation}
\ddot{q}_{n}+E_{0}^{2}q_{n}=E_{0}^{2}\frac{\alpha}{2}\left(q_{n+1}+q_{n-1}\right)\end{equation}
yield (using $E_{0}=\frac{\nu}{\sqrt{1-\alpha}}$) the dispersion
relation \begin{equation}
\omega_{k}=E_{0}\sqrt{1-\alpha\cos\theta_{k}}=\upsilon\sqrt{\frac{1-\alpha\cos\theta_{k}}{1-\alpha}}\end{equation}
and the solutions\begin{equation}
q_{n}\left(t\right)=\frac{1}{\sqrt{N}}\overset{N-1}{\underset{k=0}{\sum}}\frac{1}{\sqrt{2\omega_{k}}}\left(a_{k}e^{i\left(\theta_{k}n-\omega_{k}t\right)}+a_{k}^{\dagger}e^{-i\left(\theta_{k}n-\omega_{k}t\right)}\right)\label{eq:qn}\end{equation}
\begin{equation}
p_{n}\left(t\right)=\frac{-i}{\sqrt{N}}\overset{N-1}{\underset{k=0}{\sum}}\sqrt{\frac{\omega_{k}}{2}}\left(a_{k}e^{i\left(\theta_{k}n-\omega_{k}t\right)}-a_{k}^{\dagger}e^{-i\left(\theta_{k}n-\omega_{k}t\right)}\right)\end{equation}
where\begin{equation}
\theta_{k}=\frac{2\pi k}{N}\end{equation}

\subsubsection{The continuum limit}

Let us take the limit $N\longrightarrow\infty$, keeping the chain's
length $L$ constant: $L=Nl=const.$

From the definition of $\alpha$, one gets\begin{equation}
\frac{1}{\alpha}=1+\frac{M\nu^{2}}{2K}\end{equation}
Dimensional analysis shows us that the ratio $M/K$ must be in units
of $[Time]^{2}$. Therefore there exists a constant $\widetilde{c}$,
with velocity units, which satisfies\begin{equation}
\widetilde{c}=l\sqrt{\frac{K}{M}}\label{eq:c}\end{equation}
it will be shown that this is the typical propagation velocity of
the system. When taking the continuum limit, we shall also keep $\widetilde{c}$ constant.

Now we get\begin{equation}
\frac{1}{\alpha}=1+\frac{l^{2}\nu^{2}}{2\widetilde{c}^{2}}\end{equation}
for small enough $l$ ($l\ll1$, or equivalently $L\ll N$).\begin{equation}
\alpha=1-\frac{l^{2}\nu^{2}}{2\widetilde{c}^{2}}\end{equation}
That gives the relation between $N$ and $\alpha$, which is dependent
on the velocity $\widetilde{c}$:\begin{equation}
\alpha=1-\frac{L^{2}\nu^{2}}{2N^{2}\widetilde{c}^{2}}\end{equation}

Therefore, for $k\ll N$,\begin{equation}
\omega_{k}\approx\frac{\widetilde{c}}{L}\sqrt{4\pi^{2}k^{2}+\frac{\nu^{2}L^{2}}{\widetilde{c}^{2}}}\end{equation}
one can take $\sqrt{4\pi^{2}k^{2}+\frac{\nu^{2}L^{2}}{\widetilde{c}^{2}}}\approx2\pi k$
for $\nu L\sim\widetilde{c}$ (or smaller)
\footnote{Define the function $g\left(k\right)=\frac{2\pi k}{\sqrt{1+4\pi^{2}k^{2}}}$.
$\underset{k\longrightarrow\infty}{\lim}g\left(k\right)\rightarrow1$.
It can be seen that for $k=1$, the smallest interesting value of
$k$ (it will be shown why $k=0$ is not of interest), $g\left(1\right)\approx0.987$,
so the convergence to 1 is very fast and the approximation is good
even for small $k$'s, as wanted.
}.

Hence, for small positive $k$'s,\begin{equation}
\omega_{k}\approx2\pi k\frac{\widetilde{c}}{L}\end{equation}

\subsubsection{Interaction with the spins and causality}

Denote the two oscillators which carry the internal spin as
$A$ and $B$ (taking $A$ and $B$ as the oscillators'
indexes), and define $R=B-A$ as the angular distance, is units of
$\frac{2\pi}{N}$. Working again in the interaction picture, one gets
the functions defined in (\ref{Fa},\ref{Fc}):\begin{equation}
F_{a}\left(\tau\right)\equiv\frac{1}{N}\underset{k}{\sum}\frac{\cos\left(\omega_{k}\tau-\theta_{k}R\right)}{\omega_{k}}\label{eq:fa}\end{equation}
\begin{equation}
F_{c}\left(\tau\right)\equiv\frac{1}{N}\underset{k}{\sum}\frac{\sin\left(\omega_{k}\tau-\theta_{k}R\right)}{\omega_{k}}\label{eq:fc}\end{equation}

By definition, $\omega_{k}=\omega_{N-k}$, and therefore\begin{equation}
NF_{c}\left(\tau,R\right)=\frac{1}{\omega_{0}}\sin\left(\omega_{0}\tau\right)+\begin{cases}
\overset{\frac{N-1}{2}}{\underset{k=1}{\sum}}\frac{\sin\left(\omega_{k}\tau-\theta_{k}R\right)+\sin\left(\omega_{k}\tau+\theta_{k}R\right)}{\omega_{k}} & odd\, N\\
\overset{\frac{N-2}{2}}{\underset{k=1}{\sum}}\frac{\sin\left(\omega_{k}\tau-\theta_{k}R\right)+\sin\left(\omega_{k}\tau+\theta_{k}R\right)}{\omega_{k}}+\frac{\sin\left(\omega_{N/2}\tau-\theta_{N/2}R\right)}{\omega_{N/2}} & even\, N\end{cases}\end{equation}
and similarly\begin{equation}
NF_{a}\left(\tau,R\right)=\frac{1}{\omega_{0}}\cos\left(\omega_{0}\tau\right)+\begin{cases}
\overset{\frac{N-1}{2}}{\underset{k=1}{\sum}}\frac{\cos\left(\omega_{k}\tau-\theta_{k}R\right)+\cos\left(\omega_{k}\tau+\theta_{k}R\right)}{\omega_{k}} & odd\, N\\
\overset{\frac{N-2}{2}}{\underset{k=1}{\sum}}\frac{\cos\left(\omega_{k}\tau-\theta_{k}R\right)+\cos\left(\omega_{k}\tau+\theta_{k}R\right)}{\omega_{k}}+\frac{\cos\left(\omega_{N/2}\tau-\theta_{N/2}R\right)}{\omega_{N/2}} & even\, N\end{cases}\end{equation}

Let us analyze the time behaviour of the functions for a fixed $R$;
The sums consist of trigonometric functions with coprime frequencies,
and therefore mostly do not add to some major contribution. However,
there are special times in which that does happen.

It is well known that the extremal points and zeros of $\sin u$ and
$\cos u$ are where $u=\frac{n\pi}{2}$, for $n\in\mathbb{Z}$. Here
it happens in two cases: $\omega_{k}\tau-\theta_{k}R=\frac{n\pi}{2}$
and $\omega_{k}\tau+\theta_{k}R=\frac{n\pi}{2}$. For the first case,
the "special times" are\begin{equation}
\tau_{k}^{\left(n\right)}=\frac{R}{v_{k}}+\frac{n\pi}{2\omega_{k}}\end{equation}
where the phase velocity of each mode is defined as\begin{equation}
v_{k}=\frac{\omega_{k}}{\theta_{k}}\end{equation}
using the small $k$ approximation,\begin{equation}
v_{k}\approx\frac{N}{2\pi k}2\pi k\frac{\widetilde{c}}{L}=N\frac{\widetilde{c}}{L}\end{equation}
therefore, for small $k$'s, the phase velocity is not mode-dependent,
although the frequency is. Therefore these ``special times'' are very
near (almost equal) for small $k$'s and $n=0$.

Looking at the zeros of each argument - times at which $\omega_{k}\tau-\theta_{k}R=0$,
one gets for small positive $k$'s: \begin{equation}
\tau_{k}=\frac{R}{v_{k}}\approx\frac{LR}{\widetilde{c}N}=\frac{x}{\widetilde{c}}\end{equation}
where $x=lR$ is the "real" distance along the chain, and hence
$\widetilde{c}$ can be identified as the typical (and maximal) propagation
speed along the chain, $c$:\begin{equation}
c=\widetilde{c}=\frac{L}{N}\sqrt{\frac{K}{M}}\end{equation}

The result is that several sine functions are equal to zero in very close
times. Since in these zeros the arguments equal zero as well (turns from negative
to positive, $n=0$) - each of these sines is increasing, and hence
they add up to a significant rise in $F_{c}$, around what is expected
to be the "causal" time. This rise is not sharp, since the sum
is not infinite; But it is significant. At higher $k$'s, for which
the approximation does not apply, the arguments are zero at later
times, but their contribution is less significant, because the zero
points are less dense there (in time) and the functions are "weighted"
with a $\frac{1}{\omega_{k}}$ factor, getting smaller for larger
values of $k$.

The second significant rise occurs when $\omega_{k}\tau+\theta_{k}R=0$,
again, in almost the same time for the small $k$ modes. This time
corresponds to the $N-R$ distance (since $\sin\left(\omega_{k}\tau+\theta_{k}R\right)=\sin\left(\omega_{k}\tau-\theta_{k}\left(N-R\right)\right)$),
which, remembering the periodic nature of the chain, represents the
propagation in the other direction.

What happens to $F_{a}\left(\tau,R\right)$ at these times? When the
cosine's argument is zero, it gets its maximal value, and it is symmetric
in both the sides of the zero point: therefore, we can conclude that
$F_{a}\left(\tau,R\right)$ is almost symmetric for short times before
and after the "causal" time! Therefore, the rise in the commutator
"means nothing" for the anticommutator.

Moreover, \begin{equation}
F_{c}\left(\mbox{\ensuremath{0,R}}\right)=0-\underset{k=1}{\overset{N-1}{\sum}}\frac{1}{\omega_{k}}\sin\left(\theta_{k}R\right)=0\end{equation}
(since $1/\omega_{k}$ is symmetric around $k\sim N/2$ and $\sin\left(\theta_{k}R\right)$
is antisymmetric), as expected. So if $F_{c}\left(\mbox{\ensuremath{0,R}}\right)=0$
and there is no significant rise (or fall) until $\tau\sim\frac{x}{c}$,
it would be reasonable that for earlier times the commutator is very
small.

\begin{figure}[H]
\begin{centering}
\includegraphics[scale=0.6]{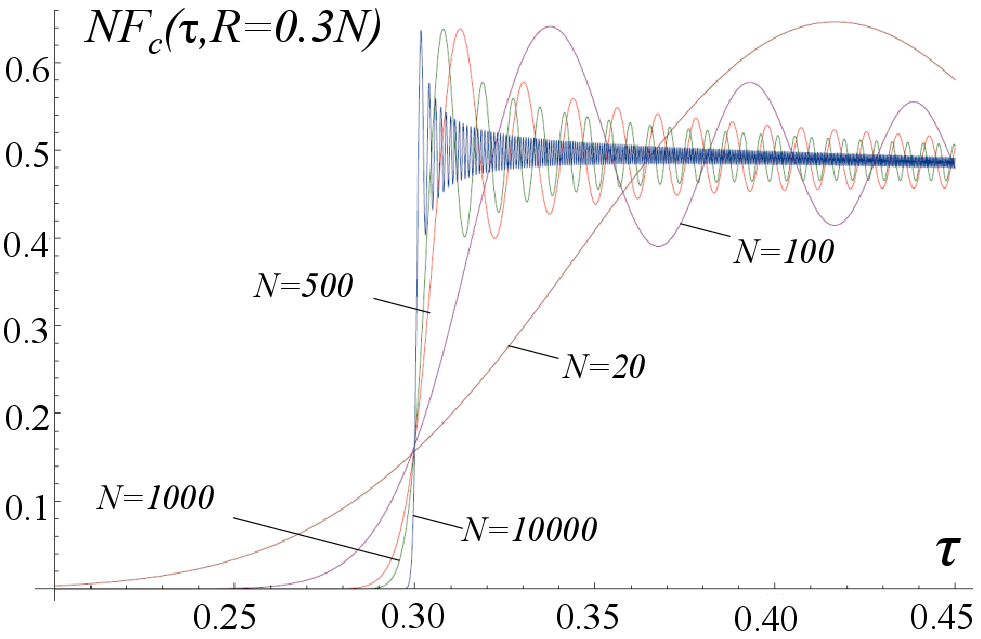}
\includegraphics[scale=0.6]{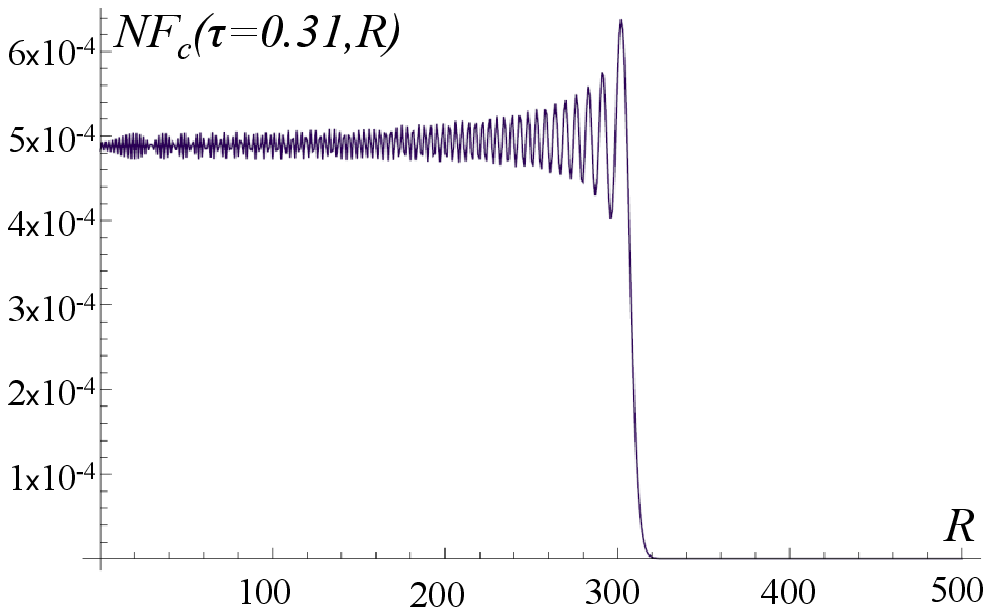}
\par\end{centering}

\centering{}\caption{On the left - Generation of a 'lightcone' - graphs of $NF_{c}\left(\tau,R=0.3N\right)$ (as defined in (\ref{eq:fc})),
for $L=1,\nu=1,c=1$; on the right - the commutator as a function
of the distance $R$ for $N=1000,\tau=0.31,L=1,\nu=1,c=1$. }
\label{Flo:Figure1}
\end{figure}

This is analogous to a lightcone, but This does not imply {}exact
causality, as in the continuum, since the commutator is not proportional
to a step function $\theta\left(t-\frac{x}{c}\right)$. However, it
does show us that the commutator's contribution is insignificant comparing
to the anticommutator's, for times "outside the lightcone",
and hence the effect does not occur because of the nonvanishing commutator,
or, in other words, the discrete nature of the system does not change
the effect. An example for the behaviour of these "causality functions"
is given in Figure \ref{Flo:Figure1}. It can be well seen from the
graphs that the "lightcone" gets sharper as $N$ grows - for
smaller $N$'s, the rise time is longer and hence the times for which
the commutator is negligible are shorter than $\frac{R}{c}$. This
is no surprise, and is due to the fact that as $N$ grows the continuum
limit, which is fully causal, is better approximated.

Before turning to the example calculations, one should note that similar
results should occur for discrete chains with other interactions,
as long as the systems have translational invariance (and therefore
the solution (\ref{eq:qn}) is applicable, and one gets the functions
(\ref{eq:fa}),(\ref{eq:fc})), and their dispersion relation is approximately
linear for small $k$'s.

\subsubsection{Example Calculations - Bare States}

For example, fix the parameters $N=100,L=1,\Omega_{A}=\Omega_{B}=2,c=1,\nu=1$.

\begin{figure}[H]
\centering{}\includegraphics[scale=0.8]{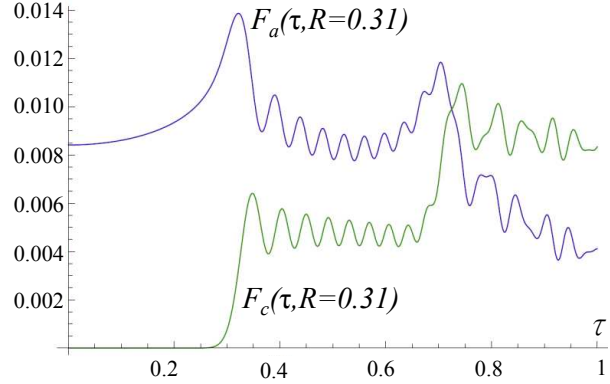}\caption{The magnitude of the commutator and the anticommutator as functions of the time, as defined in (\ref{eq:fa},\ref{eq:fc}), for $N=100,R=31,L=1,\nu=1,c=1$ }
\label{Flo:Figure2}
\end{figure}

The related commutator and anticommutator functions are plotted in
figure \ref{Flo:Figure2}.

As a first example, we choose $f_{A}\left(t\right)=f_{B}\left(t\right)\equiv1$.
Although this case forces the use of dressed states and does not allow
the post selection of all three systems, we show these naive results
as if these things were possible, in order to get the general (mathematical)
picture.

\begin{figure}[H]
\begin{centering}
\includegraphics[scale=0.6]{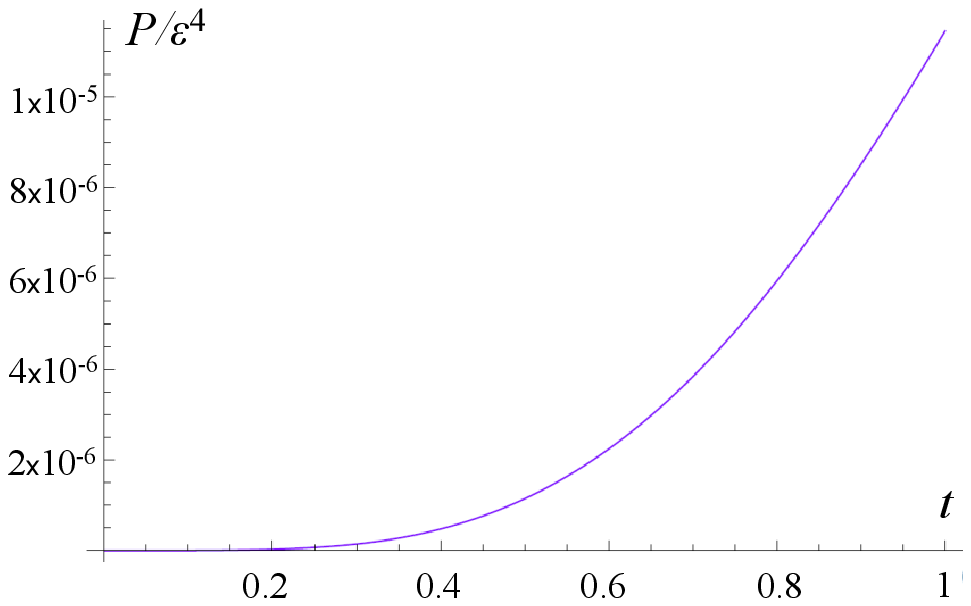}\includegraphics[scale=0.6]{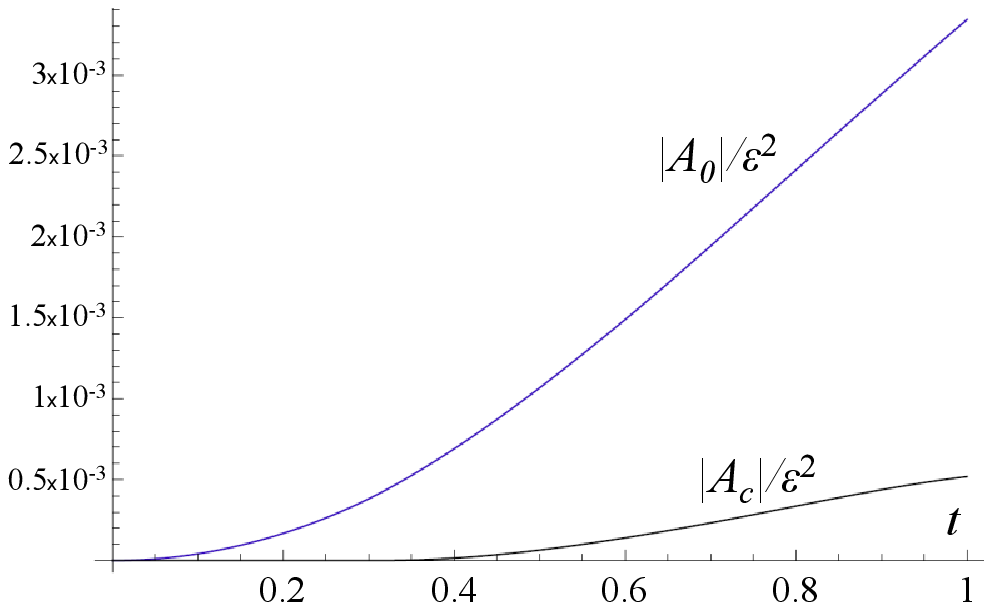}
\par\end{centering}

\centering{}\caption{The probability (on the left) and the magnitude of the separate contributions to the amplitude
(on the right) of the transition $\left|\uparrow_{A}\downarrow_{B}0_{F}\right\rangle \longrightarrow\left|\downarrow_{A}\uparrow_{B}0_{F}\right\rangle $
as a function of time, for $N=100,L=1,\Omega_{A}=\Omega_{B}=2,c=1,\nu=1$,$f_{A}\left(t\right)=f_{B}\left(t\right)\equiv1$,$R=31$.}
\label{Flo:Figure3}
\end{figure}

The results presented in Figure \ref{Flo:Figure3} are for two oscillators
with a separation of $R=31$.

\begin{figure}[H]
\centering{}\includegraphics[scale=0.8]{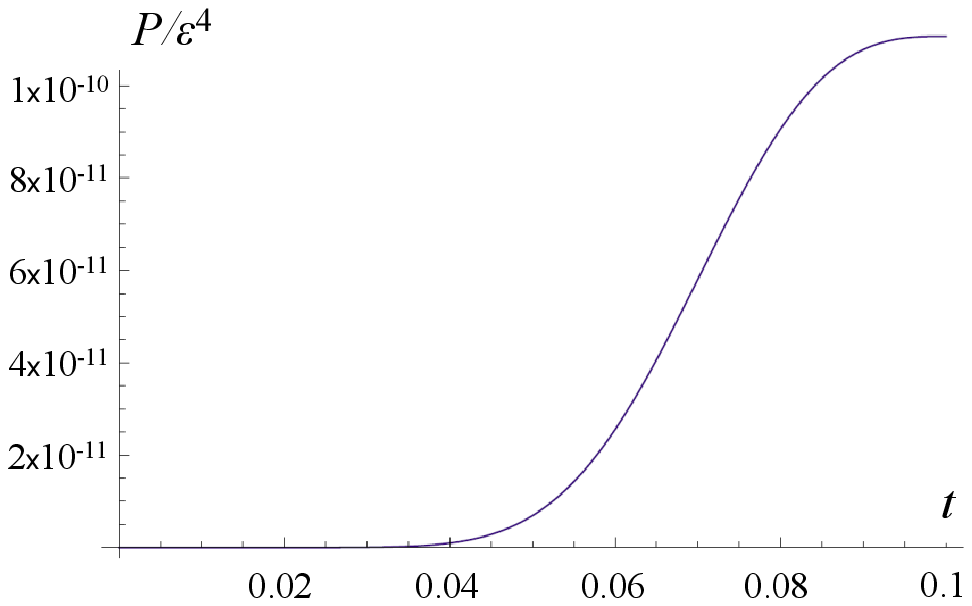}\caption{Graphic Results, showing the total probability of the transition $\left|\uparrow_{A}\downarrow_{B}0_{F}\right\rangle \longrightarrow\left|\downarrow_{A}\uparrow_{B}0_{F}\right\rangle $,
for the case of time-limited interactions: $N=100,L=1,\Omega_{A}=\Omega_{B}=2,c=1,\nu=1$,$f_{A}\left(t\right)=f_{B}\left(t\right)=\mbox{sin}^{2}\left(\frac{\pi t}{T}\right)\theta\left(t\right)\theta\left(T-t\right)$,$T=0.1,R=31$. In this case, the commutator's contribution $A_{c}$ is highly negligible comparing to $A_{0}$.}
\label{Flo:Figure4}
\end{figure}

Next we shall show another example (Figure \ref{Flo:Figure4}), the
"physical" one, with the same parameters, except for the interaction
functions, which are $f_{A}\left(t\right)=f_{B}\left(t\right)=\mbox{sin}^{2}\left(\frac{\pi t}{T}\right)\theta\left(t\right)\theta\left(T-t\right)$
: now the interactions are limited in time, so if we take $T$ short
enough, dressing effects do not have to be considered - and therefore
this is a more exact result than the previous example. In the example
shown in the following figure, $T=0.1$ which is smaller than the
"causal time" $T\sim0.31$. Moreover, turning the interactions
off allows us to perform the "nonlocal measurement" afterwards.
The results are shown in figure \ref{Flo:Figure4}.

\subsubsection{Example Calculations - Dressed States}

In order to get the dressing effects, the interaction functions were
chosen to be\begin{equation}
f_{A}\left(t\right)=f_{B}\left(t\right)=\begin{cases}
e^{\frac{t}{\tau}} & t<0\\
f_{0}\left(t\right)\equiv\cos^{2}\left(\frac{\pi t}{2T}\right)\theta\left(t\right)\theta\left(T-t\right) & t\geq0\end{cases}\end{equation}
for a very large $\tau$. It corresponds to starting with a dressed
ground state at $t=0$, and then switching the interaction off after
a duration $T$.

Plugging the harmonic chain's eigenvectors into (\ref{eq:dressed}),
we get that the time independent part of the amplitude for harmonic
chains is\begin{equation}
\frac{1}{\epsilon^{2}}G=\underset{k}{\sum}\frac{e^{i\theta_{k}\left(B-A\right)}+e^{-i\theta_{k}\left(B-A\right)}}{4N\Omega\omega_{k}\left(\Omega+\omega_{k}\right)}\end{equation}
defining, as usual, the angular distance $R=B-A$,\begin{equation}
\frac{1}{\epsilon^{2}}G=\underset{k}{\sum}\frac{\cos\left(\theta_{k}R\right)}{2N\Omega\omega_{k}\left(\Omega+\omega_{k}\right)}\end{equation}
this is, of course, a dressing effect. If the function $G\left(R\right)$
for a given chain with $N$ oscillators is examined, it can be seen
that it is a decreasing function: the "zero-time" amplitude
is bigger for closer oscillators. The minimal value, however, is not
zero: dressing occurs for every distance $R$.

\begin{figure}[H]
\centering{}\includegraphics[scale=0.6]{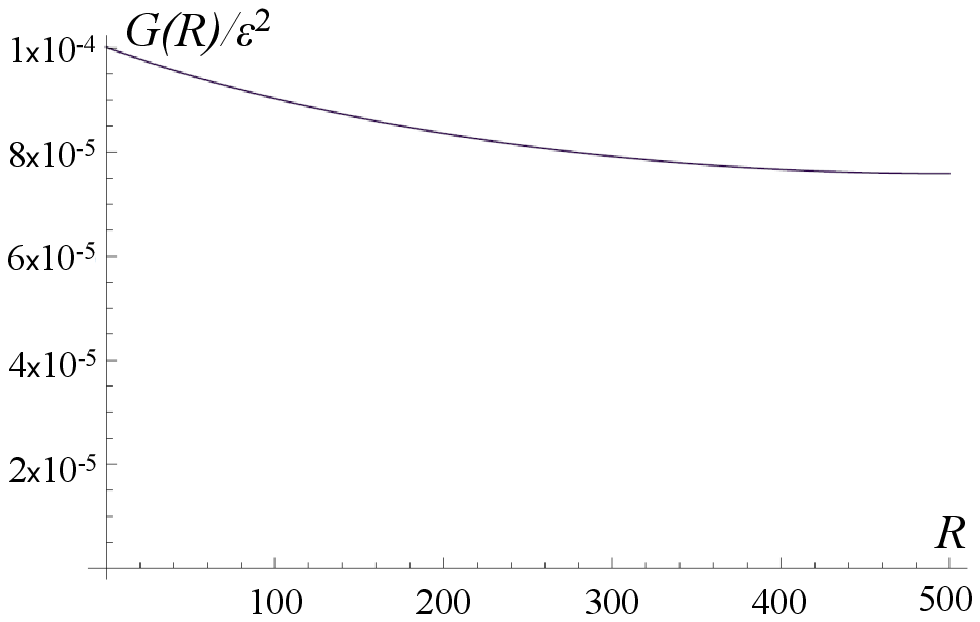}\includegraphics[scale=0.6]{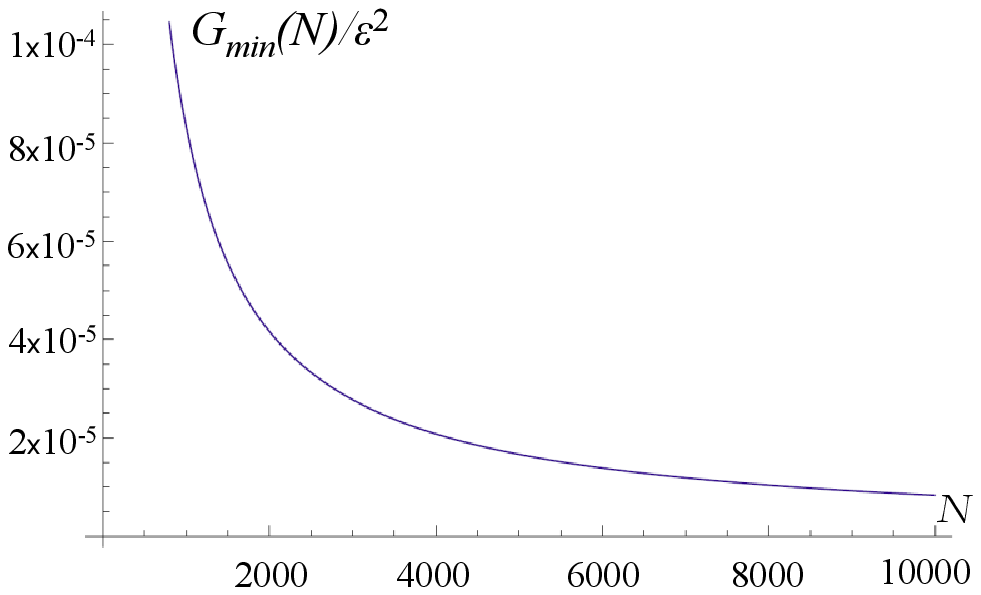}\caption{$\frac{1}{\epsilon^{2}}G$ as a function of $R$, for $N=1000$ (on
the left), and $\frac{1}{\epsilon^{2}}G_{min}\left(N\right)$ (on
the right), for even values of $N$. $L=1,\Omega=2,c=1,\nu=1$.}
\label{Flo:Figure5}
\end{figure}

The maximal distance is reached when the oscillators are on antipodal
sites on the ring: i.e., $R_{max}=\frac{N}{2}$, for a chain of an
even $N$. Therefore one can define
\begin{equation}
\frac{1}{\epsilon^{2}}G_{min}\left(N\right)=
\underset{k}{\sum}\frac{\cos\left(\theta_{k}R_{max}\right)}{2N\Omega\omega_{k}
\left(\Omega+\omega_{k}\right)}
\end{equation}
noting that
\begin{equation}
\theta_{k}R_{max}=\frac{2\pi k}{N}\frac{N}{2}=\pi k\end{equation}
one gets\begin{equation}
\frac{1}{\epsilon^{2}}G_{min}\left(N\right)=
\underset{k}{\sum}\frac{\left(-1\right)^{k}}{2N\Omega\omega_{k}
\left(\Omega+\omega_{k}\right)}\end{equation}
(for even $N$). As one can see in Figure \ref{Flo:Figure5}, this
function decreases rapidly as the chain gets longer.

Using (\ref{eq:ADressed},\ref{eq:FDressed},\ref{eq:FDressed2}), the total amplitude
is given by \begin{equation}
A\left(t\right)=\frac{\epsilon^{2}}{2N}\underset{k}{\sum\frac{1}{\omega_{k}}}\left(e^{-i\theta_{k}R}F_{1k}\left(t\right)+e^{i\theta_{k}R}F_{2k}\left(t\right)\right)\end{equation}

\begin{figure}[H]
\centering{}\includegraphics[scale=0.8]{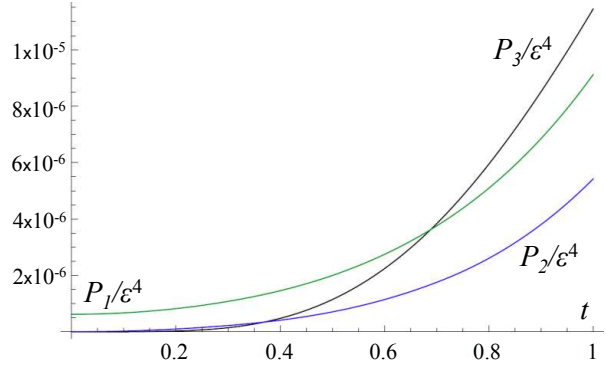}\caption{Graphic Results, showing the total transition probability in the three
dressing cases, for $N=100,L=1,\Omega=2,c=1,\nu=1,R=31,f_{0}\left(t\right)\equiv1$. ($P_{1}$ - the initial dressed state is generated by operating with $\sigma_{x}^{A}$ on the dressed ground state; $P_{2}$ - with $\sigma_{+}^{A}$; $P_{3}$ - Bare initial state, no dressing at all.)}
\label{Flo:Figure6}
\end{figure}

\begin{figure}[H]
\centering{}\includegraphics[scale=0.6]{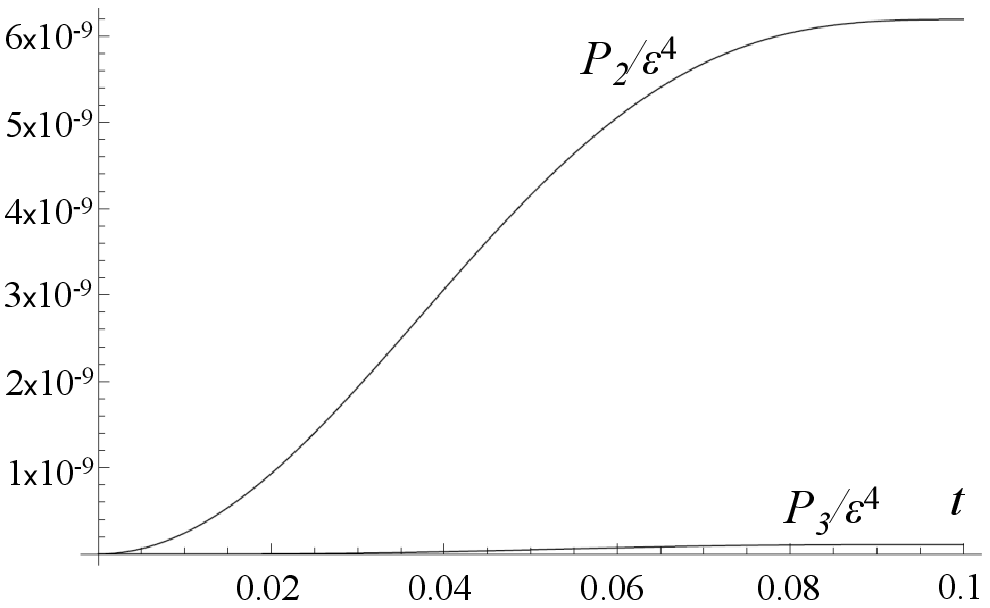}
\includegraphics[scale=0.6]{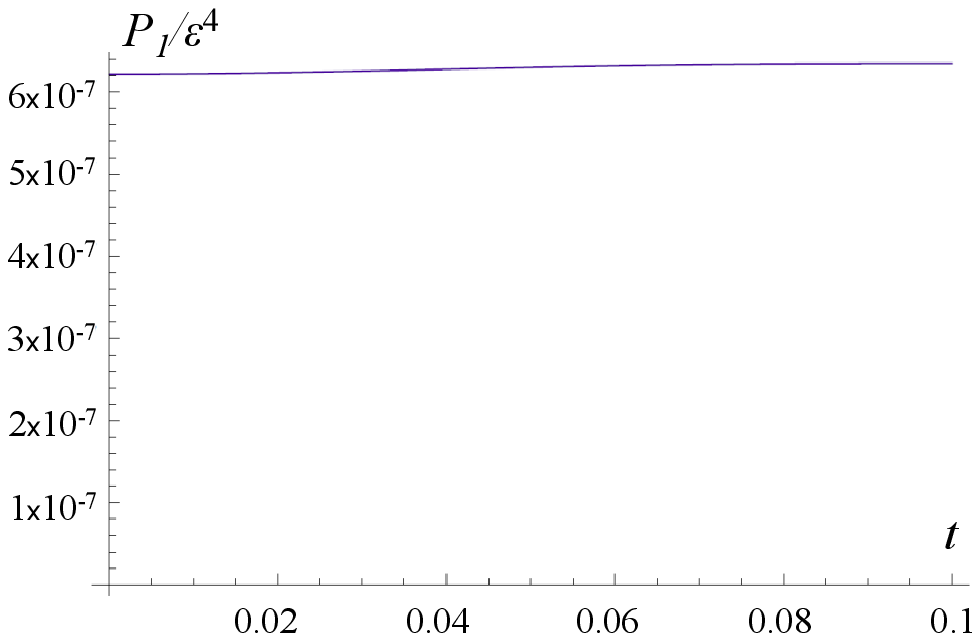}\caption{Graphic Results, showing the total transition probability in the three
dressing cases, for $N=100,L=1,\Omega=2,c=1,\nu=1,R=31,f_{0}\left(t\right)=\cos^{2}\left(\frac{\pi t}{2T}\right)\theta\left(t\right)\theta\left(T-t\right),T=0.1$. ($P_{1}$ - the initial dressed state is generated by operating with $\sigma_{x}^{A}$ on the dressed ground state; $P_{2}$ - with $\sigma_{+}^{A}$; $P_{3}$ - Bare initial state, no dressing at all.) It can be seen that the probability in the first dressing scheme is much larger than the other two; This is no surprise since the desired final state is included in the initial dressed superposition. As mentioned, the "causal time" is $T=0.31$, and it can be seen that since this dressing (constant) contribution dominates the amplitude before that time, the probability is almost constant.}
\label{Flo:Figure7}
\end{figure}

We show two example calculations - as before, the first one is
for general impression and has $f_{0}\left(t\right)\equiv1$ (Figure
\ref{Flo:Figure6}) and the second one, which is more relevant, has
$f_{0}\left(t\right)=\cos^{2}\left(\frac{\pi t}{2T}\right)\theta\left(t\right)\theta\left(T-t\right)$
(Figure \ref{Flo:Figure7}). In both the examples, $N=100,L=1,\Omega=2,c=1,\nu=1$,$R=31$, and hence the "causal time" is $T=0.31$. In the second case, $T=0.1$, which is smaller than the causal time, as expected.
One can see that in the "real" interaction, where the interactions are switched off (Figure \ref{Flo:Figure7}), the measured probabilities (i.e. in the end of the interaction time) of the dressed processes are much larger than in the bare ones; Moreover, when the first dressing scheme (with $\sigma_{x}^{A}$) is used, the probability is higher, in our example, in about two orders of magnitude than in the second scheme ($\sigma_{+}^{A}$), and is much more than in the bare case. For such a short time the constant contribution to the probability, resulting from the appearance of the desired final state in the initial dressed superposition, dominates, and hence, as can be seen in the figure, the probability of the first dressing scheme is almost constant in time.

\subsection{Ion Trap}

In this section we study the Fermi Problem for trapped ions.

Consider a linear ion trap, consisting of $N$ identical ions in a Linear Paul trap.
The
ions' free linearized Hamiltonian, $H_{F}$  amounts to a set of free harmonic oscillators (as in (\ref{eq:HF}));
 The displacement
(coordinate) operator of each ion is \begin{equation}
q_{n}=\underset{k}{\sum}\frac{D_{nk}}{\sqrt{2\omega_{k}}}\left(a_{k}+a_{k}^{\dagger}\right)\end{equation}
where $\omega_{k},a_{k},a_{k}^{\dagger}$ are as usual, and $D_{nk}$,
elements of the eigenmode vectors, are calculated numerically (as
well as the frequencies $\omega_{k}$).
In the Lamb-Dicke limit, using rotating wave approximation
for the internal degrees of freedom only and moving to the rotating frame, we have \begin{equation}
H'=-\frac{1}{2}\delta\left(\sigma_{z}^{A}+\sigma_{z}^{B}\right)+H_{F}+\epsilon\underset{n\in\left\{ A,B\right\} }{\sum}f_{n}\left(t\right)\sigma_{x}^{n}q_{n}\end{equation}
where $\delta=\omega_{L}-\Omega$ is the detuning from the laser frequency.
thus one gets, in interaction picture\begin{equation}
H'_{int}=\epsilon\underset{n\in\left\{ A,B\right\} }{\sum}f_{n}\left(t\right)\left(\sigma_{+}^{n}e^{-i\delta t}+\sigma_{-}^{n}e^{i\delta t}\right)q_{n}\left(t\right)\end{equation}
which is similar to the general interaction Hamiltonian presented
before for discrete systems: one has only to replace $\Omega$ with
$-\delta$.

The perturbative calculation turns out to be very similar to the harmonic chain case.
We therefore proceed with a nonperturbative example.

\subsubsection{A Non Perturbative Example}

As seen before, although the effect exists, it  is very
small. In order to increase its magnitude, one would like to increase the coupling
constant $\epsilon$; and use instead a non perturbative approach. The effect can then be anticipated to become significantly larger.

We demonstrate this by following \cite{ionentanglement}, and considering as a "proof of principle"   an ion trap with only
two ions. Their ground state, $\left|V\right\rangle =\left|00\right\rangle $
(in the eigenmodes basis), can be written in the local number basis
as\begin{equation}
\left|V\right\rangle =Z\overset{\infty}{\underset{n=0}{\sum}}e^{-\beta n}\left|n_{A}\right\rangle \left|n_{B}\right\rangle \end{equation}
where \begin{equation}
Z=\sqrt{1-e^{-2\beta}}\end{equation}

It is a thermal state, whose "temperature" $\beta$ can be calculated,
for example, using symplectic diagonalization and Williamson modes
\cite{reznikbotero}. \begin{equation}
e^{-\beta}=\sqrt{\frac{\lambda-\frac{1}{2}}{\lambda+\frac{1}{2}}}\end{equation}
where $\lambda$ is the symplectic eigenvalue\begin{equation}
\lambda=\frac{1}{4}\left(\sqrt{\frac{\omega_{0}}{\omega_{1}}}+\sqrt{\frac{\omega_{1}}{\omega_{0}}}\right)\end{equation}
and $\omega_{0},\omega_{1}$ are the frequencies of the system's eigenmodes.

Calculating the frequencies of our system ($\omega_{1}=\sqrt{3}\omega_{0}$),
we get $\lambda\approx0.5189$, and hence $\beta\approx1.9916,e^{-\beta}\approx0.1364$.
Due to the smallness of $e^{-\beta}$, one can write, approximately,
\begin{equation}
\left|V\right\rangle \approx\left|\widetilde{V}\right\rangle =\left|0_{A}\right\rangle \left|0_{B}\right\rangle +e^{-\beta}\left|1_{A}\right\rangle \left|1_{B}\right\rangle \end{equation}

The displacement operators, in terms of the local sites creation and annihilation
operators, are given by\begin{equation}
q_{n}=\frac{1}{\sqrt{2\omega_{0}}}\left(a_{n}+a_{n}^{\dagger}\right)\end{equation}

$T$, the time during which the interactions are open (the pulse length)
must satisfy $T\ll\frac{1}{\omega_{0}}$ - shorter than the typical
propagation time along the trap, and thus it obeys the "outside
the lightcone" demand: for ion traps,\begin{equation}
F_{a}\left(\tau\right)=\underset{k}{\sum}\frac{D_{ak}D_{bk}\cos\left(\omega_{k}\tau\right)}{\omega_{k}}\end{equation}
and\begin{equation}
F_{c}\left(\tau\right)=\underset{k}{\sum}\frac{D_{ak}D_{bk}\sin\left(\omega_{k}\tau\right)}{\omega_{k}}\end{equation}
and so, for $\tau<T\ll\frac{1}{\omega_{0}}$, one gets that $F_{c}\left(\tau\right)$
is negligible comparing to $F_{a}\left(\tau\right)$.

We consider the case that the interaction coupling constant is sufficiently strong, using strong and impulsive laser pulses, such that the free evolution Hamiltonian can be neglected. One can then approximate,
$H\approx H_{I}$, and get\begin{equation}
\left|\psi\left(t\right)\right\rangle \approx e^{i\left(\alpha_{A}\sigma_{x}^{A}\left(a+a^{\dagger}\right)+\alpha_{B}\sigma_{x}^{B}\left(b+b^{\dagger}\right)\right)}\left|\uparrow_{A}\downarrow_{B}V_{F}\right\rangle \end{equation}
where $a,a^{\dagger}$ act on $A$ and $b,b^{\dagger}$ on $B$, and $\alpha_{n}=-\frac{\epsilon}{\sqrt{2\omega_{0}}}\int_{0}^{T}dtf_{n}\left(t\right).$

Working in the Schrodinger picture, the operators are time-independent.
Therefore the time evolution operator can be separated into two commuting
parts\begin{equation}
\left|\psi\left(t\right)\right\rangle \approx e^{i\alpha_{A}\sigma_{x}^{A}\left(a+a^{\dagger}\right)}e^{i\alpha_{B}\sigma_{x}^{B}\left(b+b^{\dagger}\right)}\left|\uparrow_{A}\downarrow_{B}\widetilde{V}_{F}\right\rangle \end{equation}
which can be decomposed further:\begin{equation}
e^{i\alpha_{A}\sigma_{x}^{A}\left(a+a^{\dagger}\right)}=e^{i\alpha_{A}\sigma_{x}^{A}a^{\dagger}}e^{i\alpha_{A}\sigma_{x}^{A}a}e^{-\frac{1}{2}\left[i\alpha_{A}\sigma_{x}^{A}a^{\dagger},i\alpha_{A}\sigma_{x}^{A}a\right]}=e^{-\frac{\alpha_{A}^{2}}{2}}e^{i\alpha_{A}\sigma_{x}^{A}a^{\dagger}}e^{i\alpha_{A}\sigma_{x}^{A}a}\end{equation}
and so one obtains\begin{equation}
\left|\psi\left(t\right)\right\rangle \approx e^{-\frac{\alpha_{A}^{2}+\alpha_{B}^{2}}{2}}e^{i\alpha_{A}\sigma_{x}^{A}a^{\dagger}}e^{i\alpha_{B}\sigma_{x}^{B}b^{\dagger}}e^{i\alpha_{A}\sigma_{x}^{A}a}e^{i\alpha_{B}\sigma_{x}^{B}b}\left|\uparrow_{A}\downarrow_{B}\right\rangle \left(\left|0_{A}\right\rangle \left|0_{B}\right\rangle +e^{-\beta}\left|1_{A}\right\rangle \left|1_{B}\right\rangle \right)\end{equation}
note that\begin{equation}
e^{i\alpha_{A}\sigma_{x}^{A}a}e^{i\alpha_{B}\sigma_{x}^{B}b}\left|\uparrow_{A}\downarrow_{B}\right\rangle \left|0_{A}\right\rangle \left|0_{B}\right\rangle =\left|\uparrow_{A}\downarrow_{B}\right\rangle \left|0_{A}\right\rangle \left|0_{B}\right\rangle \end{equation}
and thus\begin{multline}
e^{-\frac{\alpha_{A}^{2}+\alpha_{B}^{2}}{2}}e^{i\alpha_{A}\sigma_{x}^{A}a^{\dagger}}e^{i\alpha_{B}\sigma_{x}^{B}b^{\dagger}}e^{i\alpha_{A}\sigma_{x}^{A}a}e^{i\alpha_{B}\sigma_{x}^{B}b}\left|\uparrow_{A}\downarrow_{B}\right\rangle \left|0_{A}\right\rangle \left|0_{B}\right\rangle =\\
e^{-\frac{\alpha_{A}^{2}+\alpha_{B}^{2}}{2}}\left(1+i\alpha_{A}\sigma_{x}^{A}a^{\dagger}+\cdots\right)\left(1+i\alpha_{B}\sigma_{x}^{B}b^{\dagger}+\cdots\right)\left|\uparrow_{A}\downarrow_{B}\right\rangle \left|0_{A}\right\rangle \left|0_{B}\right\rangle =\\
-e^{-\frac{\alpha_{A}^{2}+\alpha_{B}^{2}}{2}}\alpha_{A}\alpha_{B}\left|\downarrow_{A}\uparrow_{B}\right\rangle \left|1_{A}\right\rangle \left|1_{B}\right\rangle +\left|\tilde{\psi}\right\rangle \end{multline}

Similarly,

\begin{multline*}
e^{-\frac{\alpha_{A}^{2}+\alpha_{B}^{2}}{2}}e^{i\alpha_{A}\sigma_{x}^{A}a^{\dagger}}e^{i\alpha_{B}\sigma_{x}^{B}b^{\dagger}}e^{i\alpha_{A}\sigma_{x}^{A}a}e^{i\alpha_{B}\sigma_{x}^{B}b}\left|\uparrow_{A}\downarrow_{B}\right\rangle \left|1_{A}\right\rangle \left|1_{B}\right\rangle =\\
-e^{-\frac{\alpha_{A}^{2}+\alpha_{B}^{2}}{2}}\alpha_{A}\alpha_{B}\left|\downarrow_{A}\uparrow_{B}\right\rangle \left|0_{A}\right\rangle \left|0_{B}\right\rangle +\left|\tilde{\psi'}\right\rangle \end{multline*}

where $\left\langle \psi_{f}|\tilde{\psi}\right\rangle =\left\langle \psi_{f}|\tilde{\psi'}\right\rangle =0$.
The Amplitude is
\begin{equation}
A\left(\alpha_{A},\alpha_{B}\right)\approx-2e^{-\frac{\alpha_{A}^{2}+\alpha_{B}^{2}}{2}}\alpha_{A}\alpha_{B}e^{-\beta}\end{equation}
and therefore\begin{equation}
P\left(\alpha_{A},\alpha_{B}\right)\approx4e^{-\alpha_{A}^{2}+\alpha_{B}^{2}}\alpha_{A}^{2}\alpha_{B}^{2}e^{-2\beta}\end{equation}
if the similar pulses are used for both the ions ($\alpha_{A}=\alpha_{B}\equiv\alpha)$,
the transition probability is\begin{equation}
P\left(\alpha\right)\approx4e^{-2\alpha^{2}}\alpha^{4}e^{-2\beta}\end{equation}

This probability is maximal for $\alpha=1$: $P\left(1\right)\approx e^{-2\beta-2}\approx0.0100819\approx1\%$
.

\section{Summary and Discussion}

We have discussed the simulations of the Fermi two-atom problem within
the context of discrete systems. The fact that in such cases the interactions
between the "atoms" and the "field" are controllable, enables
us to perform post selection on all the participating systems, and therefore
the problem could be re-examined in Fermi's original formulation \cite{fermi}.

As shown in section 2, the general calculation of the relevant transition
amplitude for a discrete system, to leading order in perturbation
theory, can be decomposed into two separate parts - one is correlation-dependent,
similar to the continuous case outside the lightcone, and the other
is commutator-dependent. The second part vanishes in the continuous
case, and hence defines a light-cone analogy for our system: times
at which the second part is negligible comparing to the first are
defined as being ``outside the lightcone'', and then the first
part dominate and the discrete effect is similar to the continuous
one. We have seen, for example, for harmonic chains, that the commutator's contribution
grows around $t\sim\frac{R}{c}$, and as the number of oscillators
grow, the discrete "lightcone" is approaching a real, continuous
lightcone.

Another thing of interest was the question of dressing. Turning the
interaction on quickly and for a short time enables us to use the
bare states as the physical states of the system. However, if the
interaction is turned on adiabatically, the states undergo a dressing
process, and the physical states are the dressed ones. The dressed
states were obtained perturbatively (as in \cite{Dressing,petrosky}).
Two different dressing schemes were considered, and in one of them
the desired final states already appears in the initial dressed superposition.
Moreover, even in the bare case the nonlocal excitation cloud arises
immediately, as discussed in appendix
2.

The effect's probability is very small. In order to obtain measurable
probabilities one could try to increase the interaction strength,
leaving the perturbative regime. We have shown, as a proof of principle,
an example for an ion trap of two ions with a very strong, impulsive
interaction, for which a probability of about 1\% was calculated.
This, however, was done only with the bare states, since our dressing
process was perturbative and hence dependent on the interactions being
weak. In an experiment, however, one could turn on strong interactions
adiabatically, causing the system to dress, and observe the influence
on the measured probability.

The field's initial state could be considered further as well: for
large $N$, it might not be possible in current day experimental technologies
to cool the system to its ground state. Hence the effect with the
field in a thermal state of a low temperature is another question
worth addressing.

In conclusion, one can see, in a discrete system for which a lightcone can be
approximated, that the probability of swapping the atoms' internal degrees of freedom appears to be noncausal
as in the continuous case. As in the continuous case, there
is a nonlocal "cloud" of field excitations which probably cause
this to happen. The discrete effects, with the benefits which result
from turning on and off the interactions, can serve as a quantum simulator
of Fermi Problem, showing the effects of dressing.

\subsection*{Acknowledgements}
This work has been supported by the Israel Science Foundation grant number 920/09, the German-Israeli
foundation, and the European Commission (PICC).

\appendix

\section*{Appendix 1: Perturbative Calculation of the Bare Amplitude}
\setcounter{section}{1}
Working in the interaction picture, we get (assuming that $\epsilon$
is small enough) that at time $t>0$,\begin{multline}
\left|\psi\left(t\right)\right\rangle =Te^{-i\int_{0}^{t}dt'H_{I}\left(t'\right)}\left|\psi\left(0\right)\right\rangle =\\
\left(1-i\int_{0}^{t}dt'H_{I}\left(t'\right)-\int_{0}^{t}dt'H_{I}\left(t'\right)\int_{0}^{t'}dt''H_{I}\left(t''\right)+O\left(\epsilon^{3}\right)\right)\left|\psi\left(0\right)\right\rangle\end{multline}

Considering the relevant (bare) initial and final states (\ref{eq:InSt},\ref{eq:FiSt}),
the amplitude we seek is: \begin{equation}
A\left(t\right)=\left\langle \downarrow_{A}\uparrow_{B}0_{F}\right|\left(1-i\int_{0}^{t}dt'H_{I}\left(t'\right)-\int_{0}^{t}dt'H_{I}\left(t'\right)\int_{0}^{t'}dt''H_{I}\left(t''\right)+O\left(\epsilon^{3}\right)\right)\left|\uparrow_{A}\downarrow_{B}0_{F}\right\rangle \end{equation}

The lowest order process contributing to this amplitude is an exchange
of a virtual phonon between the two "spins". This can be written
as\begin{equation}
A\left(t\right)=-\left\langle \downarrow_{A}\uparrow_{B}0_{F}\right|\int_{0}^{t}dt'H_{I}\left(t'\right)\int_{0}^{t'}dt''H_{I}\left(t''\right)\left|\uparrow_{A}\downarrow_{B}0_{F}\right\rangle +O\left(\epsilon^{3}\right)\end{equation}
Plugging in the interaction Hamiltonians (\ref{eq:Hint}) the amplitude
can be decomposed into the two possible orders of this process (emission
- absorption and absorption - emission)\begin{multline}
A\left(t\right)=-\left\langle \downarrow_{A}\uparrow_{B}0_{F}\right|\int_{0}^{t}dt'H_{I}^{A}\left(t'\right)\int_{0}^{t'}dt''H_{I}^{B}\left(t''\right)+\int_{0}^{t}dt'H_{I}^{B}\left(t'\right)\int_{0}^{t'}dt''H_{I}^{A}\left(t''\right)\left|\uparrow_{A}\downarrow_{B}0_{F}\right\rangle +O\left(\epsilon^{3}\right)\end{multline}

After operating on the spins' spaces, the amplitude (in lowest order)
is

\begin{multline}
-\frac{A\left(t\right)}{\epsilon^{2}}=\int_{0}^{t}dt'f_{A}\left(t'\right)e^{-i\Omega_{A}t'}\int_{0}^{t'}dt''f_{B}\left(t''\right)e^{i\Omega_{B}t''}\left\langle 0\right|q_{A}\left(t'\right)q_{B}\left(t''\right)\left|0\right\rangle +\\
\int_{0}^{t}dt''f_{B}\left(t''\right)e^{i\Omega_{B}t''}\int_{0}^{t''}dt'f_{A}\left(t'\right)e^{-i\Omega_{A}t'}\left\langle 0\right|q_{B}\left(t''\right)q_{A}\left(t'\right)\left|0\right\rangle \equiv-\frac{A_{AB}\left(t\right)+A_{BA}\left(t\right)}{\epsilon^{2}}\label{eq:}\end{multline}

Changing the order of integration and the order of operators in both
the terms $A_{AB}\left(t\right),A_{BA}\left(t\right)$ we get\begin{multline}
-\frac{A_{AB}\left(t\right)}{\epsilon^{2}}=\int_{0}^{t}dt''f_{B}\left(t''\right)e^{i\Omega_{B}t''}\int_{t''}^{t}dt'f_{A}\left(t'\right)e^{-i\Omega_{A}t'}\left\langle 0\right|q_{B}\left(t''\right)q_{A}\left(t'\right)\left|0\right\rangle \\
+\int_{0}^{t}dt''f_{B}\left(t''\right)e^{i\Omega_{B}t''}\int_{t''}^{t}dt'f_{A}\left(t'\right)e^{-i\Omega_{A}t'}\left\langle 0\right|\left[q_{A}\left(t'\right),q_{B}\left(t''\right)\right]\left|0\right\rangle \end{multline}
\begin{multline}
-\frac{A_{BA}\left(t\right)}{\epsilon^{2}}=\int_{0}^{t}dt'f_{A}\left(t'\right)e^{-i\Omega_{A}t'}\int_{t'}^{t}dt''f_{B}\left(t''\right)e^{i\Omega_{B}t''}\left\langle 0\right|q_{A}\left(t'\right)q_{B}\left(t''\right)\left|0\right\rangle \\
+\int_{0}^{t}dt'f_{A}\left(t'\right)e^{-i\Omega_{A}t'}\int_{t'}^{t}dt''f_{B}\left(t''\right)e^{i\Omega_{B}t''}\left\langle 0\right|\left[q_{B}\left(t''\right),q_{A}\left(t'\right)\right]\left|0\right\rangle \end{multline}
adding each of the changed terms to the other term unchanged, a new
expression for the amplitude is reached. Adding the two expressions
obtained that way and dividing the sum by 2, one gets (\ref{eq:Amplitude}).

\appendix
\section*{Appendix 2: Estimation of the Field Excitations Distribution}
\setcounter{section}{2}

Another thing of interest is the distributions of field excitations
("phonons") in the system, as a function of the sites. Following \cite{petrosky}, we define the excitations distribution at site
$n$, time $t$, as\begin{equation}
D_{n}\left(t\right)=\left\langle \psi_{i}\left(t\right)\right|q_{n}^{+}\left(t\right)q_{n}^{-}\left(t\right)\left|\psi_{i}\left(t\right)\right\rangle \end{equation}
(in interaction picture), where\begin{equation}
q_{n}^{+}\left(t\right)=\overset{N-1}{\underset{k=0}{\sum}}\lambda_{nk}^{*}a_{k}^{\dagger}e^{i\omega_{k}t}\end{equation}
\begin{equation}
q_{n}^{-}\left(t\right)=\overset{N-1}{\underset{k=0}{\sum}}\lambda_{nk}a_{k}e^{-i\omega_{k}t}\end{equation}
create/destroy a field excitation at site $n$.\begin{equation}
D_{n}\left(t\right)=\underset{k,l}{\sum}\lambda_{nk}^{*}\lambda_{nl}e^{i\left(\omega_{k}-\omega_{l}\right)t}\left\langle \psi_{i}\left(t\right)\right|a_{k}^{\dagger}a_{l}\left|\psi_{i}\left(t\right)\right\rangle \end{equation}
using (\ref{eq:DressedTimeEvolution}),\begin{equation}
D_{n}\left(t\right)=\epsilon^{2}\underset{k,l}{\sum}\lambda_{nk}^{*}\lambda_{nl}e^{i\left(\omega_{k}-\omega_{l}\right)t}\left(c_{Ak}^{*}\left(t\right)c_{Al}\left(t\right)+c_{Bk}^{*}\left(t\right)c_{Bl}\left(t\right)\right)+O\left(\epsilon^{4}\right)\label{eq:PhonDist}\end{equation}
where\begin{equation}
c_{Ak}\left(t\right)=\lambda_{Ak}^{*}\left(\frac{d_1}{\Omega+\omega_{k}}+i\int_{0}^{t}dt'f_{0}\left(t'\right)e^{-i\left(\Omega-\omega_{k}\right)t'}\right)\end{equation}
\begin{equation}
c_{Bk}\left(t\right)=\lambda_{Bk}^{*}\left(\frac{d_1+d_2}{\Omega+\omega_{k}}+i\int_{0}^{t}dt'f_{0}\left(t'\right)e^{i\left(\Omega+\omega_{k}\right)t'}\right)\end{equation}
$d_1,d_2$ are the dressing factors defined previously. It
can be seen, therefore, that it the initial state was bare, $D_{n}\left(t=0\right)=0$
as expected - but the ``tails'' of field excitations will be spread
out for any $t>0$!

As before, the leading order does not involve any mutual dressing
factors. In other words, in the leading order, the phonon distribution
is just an addition of the field excitation distributions of the two
sites coupled to the field. Therefore we can deduce what is the field
excitations distribution of a single site: if it is in a spin-up state
(such as $A$), \begin{equation}
D_{n}^{\uparrow}\left(t\right)=\epsilon^{2}\underset{k,l}{\sum}\lambda_{nk}^{*}\lambda_{nl}e^{i\left(\omega_{k}-\omega_{l}\right)t}c_{Ak}^{*}\left(t\right)c_{Al}\left(t\right)+O\left(\epsilon^{4}\right)\end{equation}
and if it is in a spin-down state (such as $B$), \begin{equation}
D_{n}^{\downarrow}\left(t\right)=\epsilon^{2}\underset{k,l}{\sum}\lambda_{nk}^{*}\lambda_{nl}e^{i\left(\omega_{k}-\omega_{l}\right)t}c_{Bk}^{*}\left(t\right)c_{Bl}\left(t\right)+O\left(\epsilon^{4}\right)\end{equation}
and again, for the Fermi Problem, in leading order,\begin{equation}
D_{n}\left(t\right)=D_{n}^{\uparrow}\left(t\right)+D_{n}^{\downarrow}
\left(t\right)+O\left(\epsilon^{4}\right)\end{equation}

One should note that despite the fact that $q_{n}^{+}q_{n}^{-}$ could
serve as a qualitative analogy to the local site phonon number operator,
it is a nonlocal observable since it cannot be expressed by the local
operators $q_{n},p_{n}$. Therefore it would not be possible to simulate
the results of this appendix, and it could serve only for a qualitative impression.

\subsubsection*{Example for the Harmonic Chain}

We present an example for the generation of "nonlocal cloud", when the
initial state is bare (figure B1).

\begin{figure}[H]
\begin{centering}
\includegraphics[scale=0.8]{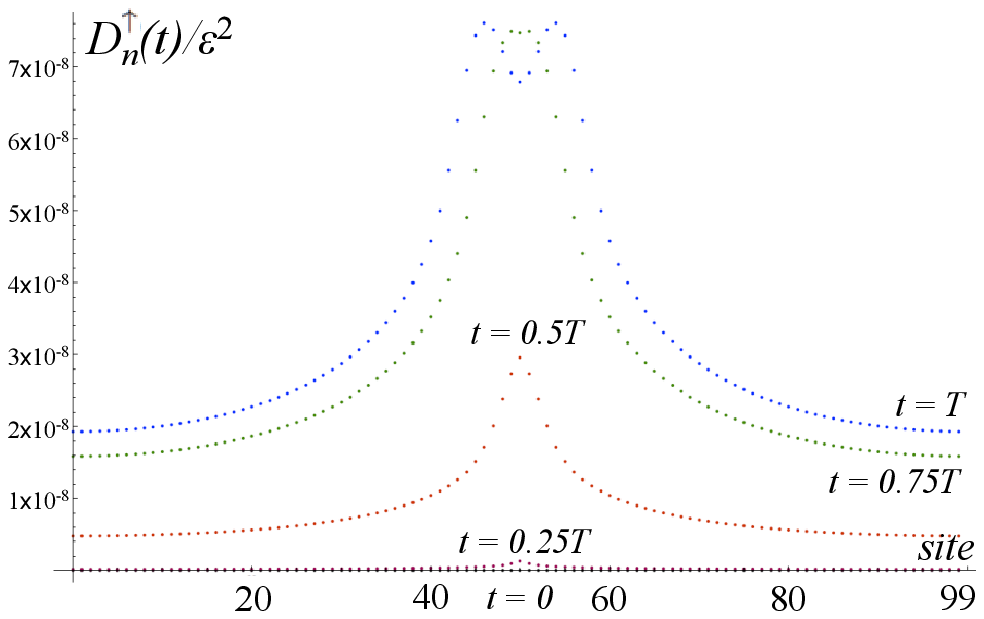}\caption{A Plot of the local field excitations distribution, for an initial
bare state, of the single spin-up atom in the 50th site of a chain
of 100 sites, with $N=100,L=1,\Omega=2,c=1,\nu=1$,$f_{A}\left(t\right)=f_{B}\left(t\right)=\mbox{sin}^{2}\left(\frac{\pi t}{T}\right)\theta\left(t\right)\theta\left(T-t\right)$,$T=0.1$.
The generation and spreading of the nonlocal "cloud" can be
seen.}

\par\end{centering}

\label{Flo:Figure8}
\end{figure}

\bibliographystyle{unsrt}

\end{document}